\documentclass[12pt]{iopart}
\usepackage{graphicx}

\usepackage{amssymb}
\usepackage{amsbsy}

\begin{document}

\title{Longitudinal and Transverse structure functions \\ 
in high Reynolds-number turbulence}
\author{Rainer Grauer$^1$, Holger Homann$^{1,2}$, Jean-Fran\c{c}ois Pinton$^3$}

\address{$^1$ Theoretische Physik I, Ruhr-Universit\"at Bochum,
Universit\"atsstr. 150, D44780 Bochum (Germany)}
\address{$^2$ Universit\'e de Nice-Sophia Antipolis, CNRS, Observatoire de la
C\^ote d'Azur, Laboratoire Cassiop\'ee, Bd. de l'Observatoire, 06300 Nice, France}
\address{$^3$ Laboratoire de Physique, UMR5672, CNRS \& \'{E}cole Normale Sup\'{e}rieure de Lyon,\\ 
46 all\'{e}e d'Italie, F69007 Lyon (France)}

\date{\today}

\begin{abstract}
  Using exact relations between velocity structure
  functions~\cite{hill:2001,hill-boratav:2001,yakhot:2001} and neglecting
  pressure contributions in a first approximation, we obtain a closed system and
  derive simple order-dependent rescaling relationships between longitudinal and transverse
  structure functions.  
  By means of numerical data with turbulent Reynolds
  numbers ranging from $\Re_\lambda=320$ to $\Re_\lambda=730$, we 
  establish a clear correspondence between their respective scaling range, 
  while confirming that their scaling exponents do differ.
  This difference does not seem to depend on Reynolds number. Making
  use of the Mellin transform, we further map longitudinal to (rescaled)
  transverse probability density functions. \\
\end{abstract}


\maketitle

\section{Introduction}
Intermittency is an ubiquitous feature of fluid turbulence: the scaling
properties of flow quantities differ from Kolmogorov's mean field
theory~\cite{frisch:1995}. For instance, in the inertial range of scales where
flow properties are assumed to be independent of the details of energy injection
and dissipation, the velocity increments do not have a monofractal structure. In
homogeneous isotropic turbulence (HIT), two directions only matter in the
computation of a velocity increment: the {\it longitudinal} one, $\Delta_r u$
taken along the separation and the {\it transverse} one, $\Delta_r v$ in which
the difference of velocities components perpendicular to the separation are
computed.  Velocity structure functions are then defined as the ensemble
average: $S_{n,m}(r) = \langle (\Delta_r u)^n (\Delta_r v)^m \rangle$. For HIT,
the von K\'arm\'an-Howarth relationship taken in the inertia range, leads to
Kolmogorov's 4/5th law: $(\Delta u_r)^3 = -\frac{4}{5} \langle \epsilon \rangle
r$, where  $\langle \epsilon \rangle$ is the mean energy dissipation rate per
unit mass. While a monofractal inertial range behavior would then lead to
$S_{n,0} \propto r^{\zeta_n}$ with $\zeta_n \sim n/3$, intermittency means that
$\zeta_n$ is a non-linear (concave) function of $n$. Numerous works have been
devoted to the study of the functional form of $\zeta_n$. We
focus on the possible link between the longitudinal and transverse
structure functions $S_{n,0}(r)$ and $S_{0,n}(r)$.  There exist theoretical
arguments that longitudinal and transverse show the same
scaling~\cite{biferale-procaccia:2005}. However, both experimental
data~\cite{noullez-wallace-etal:1997, vanderwater-herweijer:1999,
zhou-antonia:2000,shen-warhaft:2002} and numerical
simulations~\cite{boratav-pelz:1997,gotoh-fukayama-etal:2002,ishihara-gotoh-etal:2009,benzi-biferale-etal:2010}
show consistently different scaling exponents for longitudinal and transverse
structure functions. Whether this difference can be attributed to a persistent
small scale anisotropy~\cite{biferale-lanotte-etal:2008, romano-antonia:2001} or
to a finite Reynolds number effect~\cite{hill:2001} is an unsolved question
to which we will come back below. Here, we note that in the case of the
direct cascade in electron-magnetohydrodynamic~\cite{germaschewski-grauer:1999}
it was demonstrated numerically that the difference vanishes with increasing
numerical resolution and thus is a finite Reynolds number effect.

In this article, we rather  focus on the correspondence between scaling
ranges of the longitudinal and transverse structure functions.  Our approach is
based on the observation that even though the (real space) velocity field of a
turbulent flow coarse-grained at a scale $r$ is not smooth, the structure
functions are smooth (differentiable) functions of $r$.  We thus use the
structure of the Navier-Stokes equation together with assumptions to derive
constitutive relationships between $S_{n,0}(r)$ and $S_{0,n}(r)$. Specifically,
we shall neglect the contributions from the pressure term. We start with exact
scaling expressions derived by Hill~\cite{hill:2001}, Hill and
Boratav~\cite{hill-boratav:2001} and Yakhot~\cite{yakhot:2001}; we then obtain
rescaling relationships between longitudinal and transverse structure functions.
Our first finding is that, after rescaling, the longitudinal and transverse
structure functions share the {\it same} inertial range, i.e. the same width in
the extent of of scales where self-similarity is observed. This is important
because the question of the location and span of the inertial range is often an
issue in the analysis of turbulent data at (necessarily) finite Reynolds number.
We stress, however, that the value of the longitudinal and transverse scaling
exponents do differ. A second outcome of our simple ansatz is a direct mapping,
using the Mellin transform, of the transverse and longitudinal probability
density functions (PDFs). Differences which persist after the mapping are then
due to the effect of the neglected terms, as pointed in some previous attempts 
by Yakhot \cite{yakhot:2001} and Gotoh and Nakano \cite{gotoh-nakano:2003}.

\section{Rescaling relations between longitudinal and transverse structure functions}

Our calculation trace back to the observation by Siefert and
Peinke~\cite{siefert-peinke:2004} that the K\'arm\'an equation (see K\'arm\'an
and Howarth~\cite{karman-howarth:1938}) relating second order longitudinal and
transverse structure functions can be interpreted as a Taylor expansion of a
smooth function. To see this, we start with the K\'arm\'an equation
\begin{equation}
	 S_{0,2}(r) =  S_{2,0}(r) + \frac{r}{2} \frac{\partial}{\partial r} S_{2,0}(r) \ ,
	 \label{eq:karman1}
\end{equation}
which is exact, and contains no contribution from the pressure
-- it is a statement of incompressibility. Siefert and Peinke
\cite{siefert-peinke:2004} observed that the structure function is a smooth
function of $r$ and that if the scale $r$ is chosen in the inertial range
{\it i.e.} ``small'' compared to the integral scale $L$, eqn.
(\ref{eq:karman1}) can be seen as a Taylor expansion:
\begin{equation}
	S_{0,2}(r) \approx  S_{2,0}(r+\frac{r}{2}) = S_{2,0}(\frac{3}{2} r) \ ,
\end{equation}
where the function $S_{2,0}$ is expanded about $r$
for consistency with the exact relationship~(\ref{eq:karman1}). 
In \cite{siefert-peinke:2004} evidence from experimental data Taylor-based
Reynolds numbers between $180$ and $550$ was presented to support this view. The
success of the approach introduced by Siefert and Peinke
\cite{siefert-peinke:2004} motivated us to extend their reinterpretation of
differential relations to structure functions of higher orders, making use of
the exact relationships derived by Hill \& Boratav~\cite{hill-boratav:2001},
Hill~\cite{hill:2001} and Yakhot~\cite{yakhot:2001}. Hill derived these
relations directly by inventing a clever matrix algorithm which allowed him to
efficiently simplify the derivation and calculations. Yakhot~\cite{yakhot:2001},
 on the other hand, derived an equation for the characteristic function
$Z=\langle \mathbf{\lambda} \cdot \Delta \mathbf{u}_r\rangle$ where
$\Delta\mathbf{u}_r$ denotes a velocity increment over the distance $r$.
Structure function relations can then be obtained by differentiating the
characteristic function $Z$.

As an illustrative example, consider the relation for even order mixed structure
functions derived by Yakhot~\cite{yakhot:2001}:
\begin{equation}
	\frac{\partial S_{2n,0}}{\partial r} + \frac{2}{r} S_{2n,0} 
	+ \frac{2 (2n-1)}{r} S_{2n-2,2} = C_p + C_f \; .
\end{equation}
The term $C_p$ contains contributions from the (unknown) pressure field
and is the reason why the system cannot be closed. The term $C_f$ contains
contributions from the large scale forcing and can safely be ignored in the
inertial range as proposed by  Kurien and
Sreenivasan~\cite{kurien-screenivasan:2001}. These authors also analyzed and
compared these relations to measurements in atmospheric turbulence at a
Taylor-based Reynolds-number of about $10 \, 700$. One of their findings was that
for even order structure functions the pressure contributions can be an order of
magnitude smaller than the terms directly related to the structure functions.
A detailed numerical study on the role of the pressure term has been realized 
by Gotoh and Nakano~\cite{gotoh-nakano:2003}. 
In order to obtain closed expressions, we shall hereafter neglect the pressure 
contributions. Although this assumption is quite crude, we are already able to  
capture the dominant features of the relationship between longitudinal and transverse 
structure functions, such as amplitudes and common inertial range.
Pressure (or energy injection) contributions will then appear as 
departures from predictions of this closed system of equations.

In order to demonstrate the procedure, we start with formulas for the 4-th order
structure functions  and neglect contributions from the pressure and
the large scale forcing (see also eqn.(11) and (13) in~\cite{kurien-screenivasan:2001})
\begin{eqnarray}
	 3 S_{2,2}(r) &\approx S_{4,0}(r) + \frac{r}{2} \frac{\partial}{\partial r}  
	 S_{4,0}(r) &\approx S_{4,0}(\frac{3}{2} r) \ , \label{eq:s22}\\
	 \frac{1}{3} S_{0,4}(r) &\approx S_{2,2}(r) + \frac{r}{4}
	 \frac{\partial}{\partial r} S_{2,2}(r) &\approx S_{2,2}(\frac{5}{4} r) \ , \label{eq:s04}
\end{eqnarray}
which can be combined into
\begin{equation}
	S_{4,0}\left(\frac{3}{2}\frac{5}{4} r\right) \approx S_{0,4}(r) \; .
\end{equation}
For the $6$th order structure functions we get similarly (see eqn.(12), (15)
and (14) in~\cite{kurien-screenivasan:2001})
\begin{eqnarray*}
	5 S_{4,2}(r) &\approx S_{6,0}(r) + \frac{r}{2} \frac{\partial}{\partial r}
	S_{6,0}(r) &\approx S_{6,0}(\frac{3}{2} r) \ , \\
	S_{2,4}(r) &\approx S_{4,2}(r) + \frac{r}{4} \frac{\partial}{\partial r}
	S_{4,2}(r) &\approx S_{4,2}(\frac{5}{4} r) \ , \\
	\frac{1}{5} S_{0,6}(r) &\approx S_{2,4}(r) + \frac{r}{6}
	\frac{\partial}{\partial r} S_{2,4}(r) &\approx S_{2,4}(\frac{7}{6} r) \; .
\end{eqnarray*}
Again, combining these equations results in the simple relation
\begin{equation}
	S_{6,0}\left(\frac{3}{2}\frac{5}{4}\frac{7}{6} r \right) \approx S_{0,6}(r) \; .
\end{equation}
In general, the rescaling for even order structure functions reads
\begin{equation}
	S_{n,0}\left(\frac{3}{2} \frac{5}{4} \ldots \frac{n+1}{n} r\right) =
	S_{n,0}\left(\frac{\Gamma(n+2)}{2^n \Gamma^2(n/2+1)} r\right) \approx
	S_{0,n}(r)
	\label{eq:rescaling}
\end{equation}

\begin{figure*}[b]
\centerline{\includegraphics[width=0.8\columnwidth]{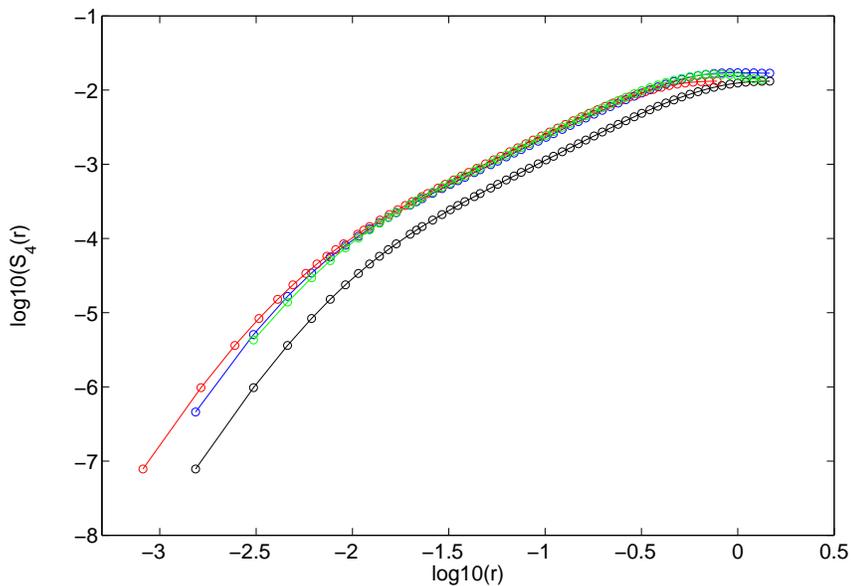}}

\caption{Comparison of structure functions: longitudinal (black), transverse (blue), rescaled longitudinal
(red) and the one using the differential relation (\ref{eq:diffrel}) (green).}
\label{fig:diffrel}
\end{figure*}

In order to demonstrate that the Taylor expansion is valid also for
higher order structure functions, we look at the differential relation for the
4th order structure function obtained from eqn. (\ref{eq:s22}) and (\ref{eq:s04})
\begin{equation}
S_{04}(r) = S_{40}(r) + \frac{7}{8} \frac{d S_{40}(r)}{dr} r 
          + \frac{1}{8} \frac{d^2 S_{40}(r)}{d r^2} r^2 \ . \label{eq:diffrel}
\end{equation}
 In Fig. \ref{fig:diffrel}
we compare the longitudinal (black), transverse (blue), rescaled longitudinal
(red) and the one using the differential relation (green). The difference
between the rescaled longitudinal and the one using the differential
relation is negligible.

In order to test our results, we use numerical data from pseudo-spectral
simulations of incompressible Navier-Stokes turbulence, as described
in~\cite{homann-kamps-etal:2009} -- the \textsc{LaTu} code. A
statistically stationary flow is maintained by keeping constant the Fourier
modes in the two lowest shells. All results are averaged over several large-eddy
turn-over times (over two in the case of $\Re_\lambda = 730$). Parameters of
these high-Reynolds number simulations  are given in
Table~\ref{table:param}. \\

\begin{table*}
  \centering
  \begin{tabular}{ccccccccccccc}
    $\Re_{\lambda}$&$u_\mathrm{rms}$& $\epsilon_\mathrm{k}$&$\nu$            & $dx$              & $\eta$           &$\tau_\eta$&$L$   &$T_L$& $N^3$ \\
    \hline 
    $730$         &$0.192$        &$3.8\cdot 10^{-3}$   &$1. \cdot 10^{-5}$&$1.53\cdot 10^{-3}$&$7.2 \cdot 10^{-4}$& 0.05      &1.85 & 9.6  &$4096^3$ \\
    $460$         &$0.189$        &$3.6\cdot 10^{-3}$   &$2.5\cdot 10^{-5}$&$3.07\cdot 10^{-3}$&$1.45\cdot 10^{-3}$& 0.083     &1.85 & 9.9  &$2048^3$ \\
    $320$         &$0.187$        &$3.5\cdot 10^{-3}$   &$0.5\cdot 10^{-4}$&$6.14\cdot 10^{-3}$&$2.45\cdot 10^{-3}$& 0.12      &1.85 & 10   &$1024^3$ \\
 \end{tabular}
 \caption{\label{table} Parameters of the numerical simulations.
    $\Re_\lambda = \sqrt{15 u_\mathrm{rms} L/ \nu}$: Taylor-Reynolds number, 
    $u_\mathrm{rms}$: root-mean-square velocity, $\epsilon_\mathrm{k}$: mean    kinetic energy dissipation rate, 
    $\nu$: kinematic viscosity, $dx$: grid-spacing, 
    $\eta =(\nu^3/\epsilon_\mathrm{k})^{1/4}$: Kolmogorov dissipation length
    scale, $\tau_\eta = (\nu/\epsilon_\mathrm{k})^{1/2}$: Kolmogorov
    time scale, $L = (2/3E)^{3/2}/\epsilon_\mathrm{k}$: integral
    scale, $T_L = L/u_\mathrm{rms}$: large-eddy turnover time, $N^3$:
    number of collocation points.}
    \label{table:param}
\end{table*}

\begin{figure*}[t]
  \centerline{
  \includegraphics[width=0.45\columnwidth]{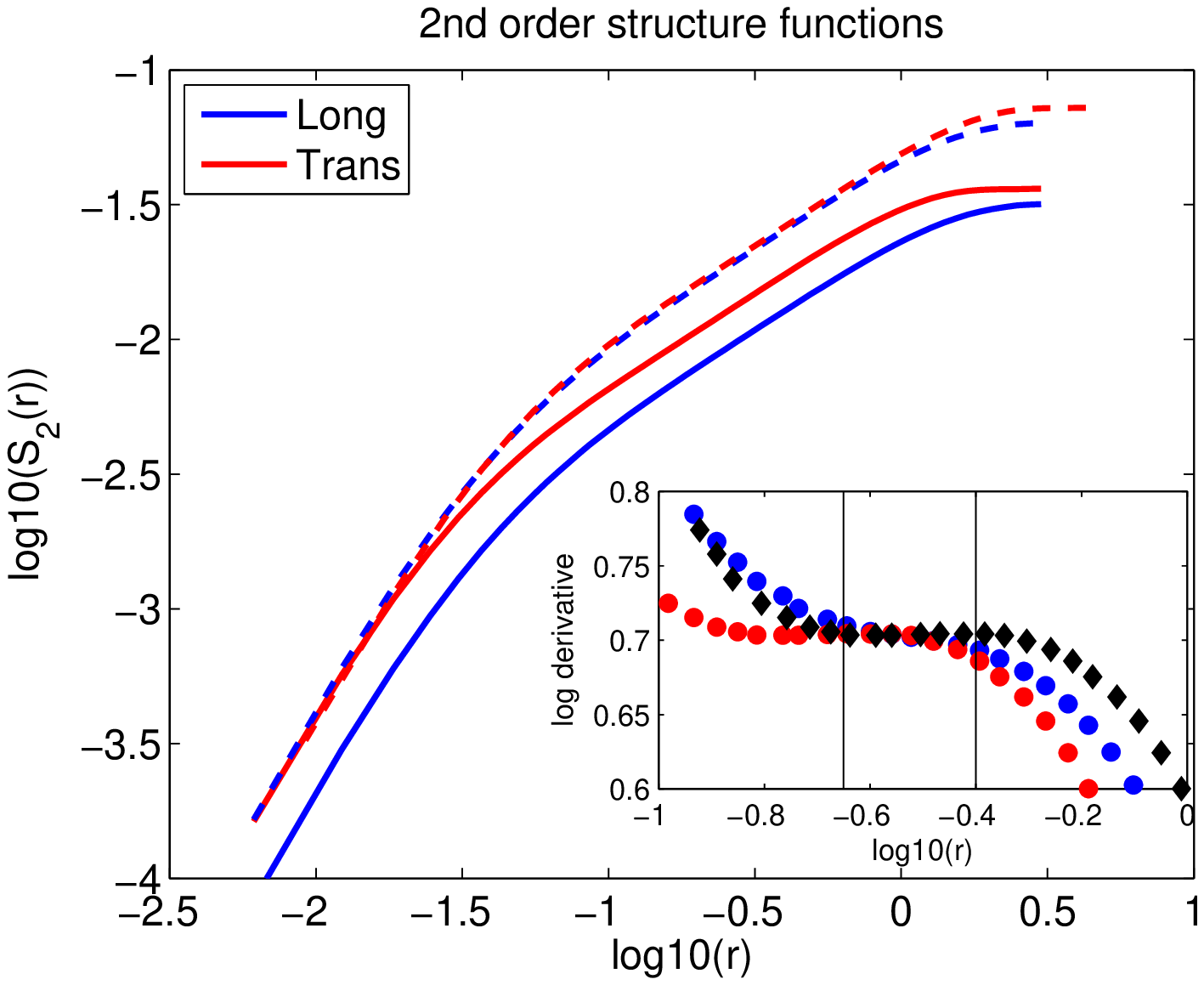} 
  \includegraphics[width=0.45\columnwidth]{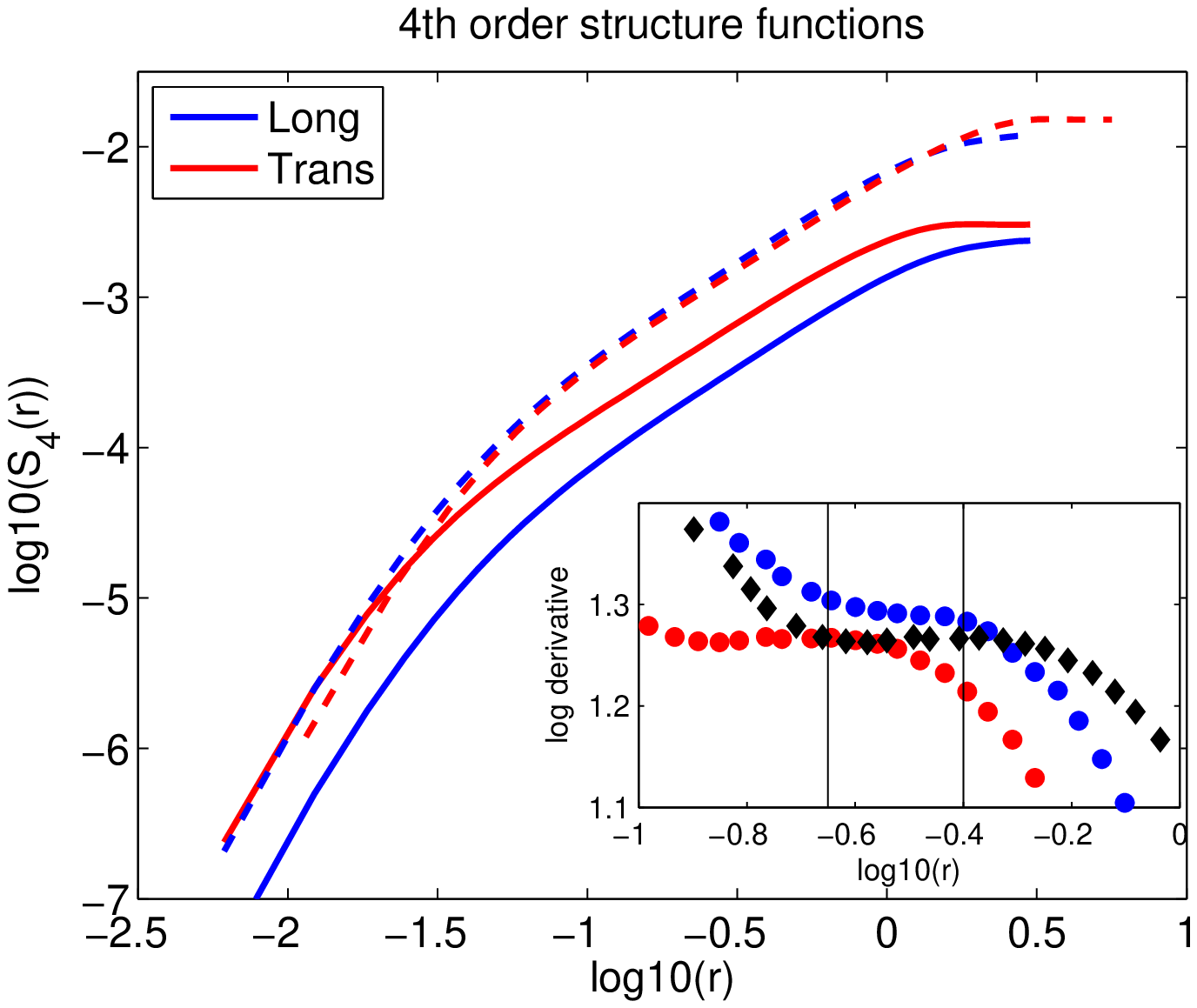}}
  \centerline{
  \includegraphics[width=0.45\columnwidth]{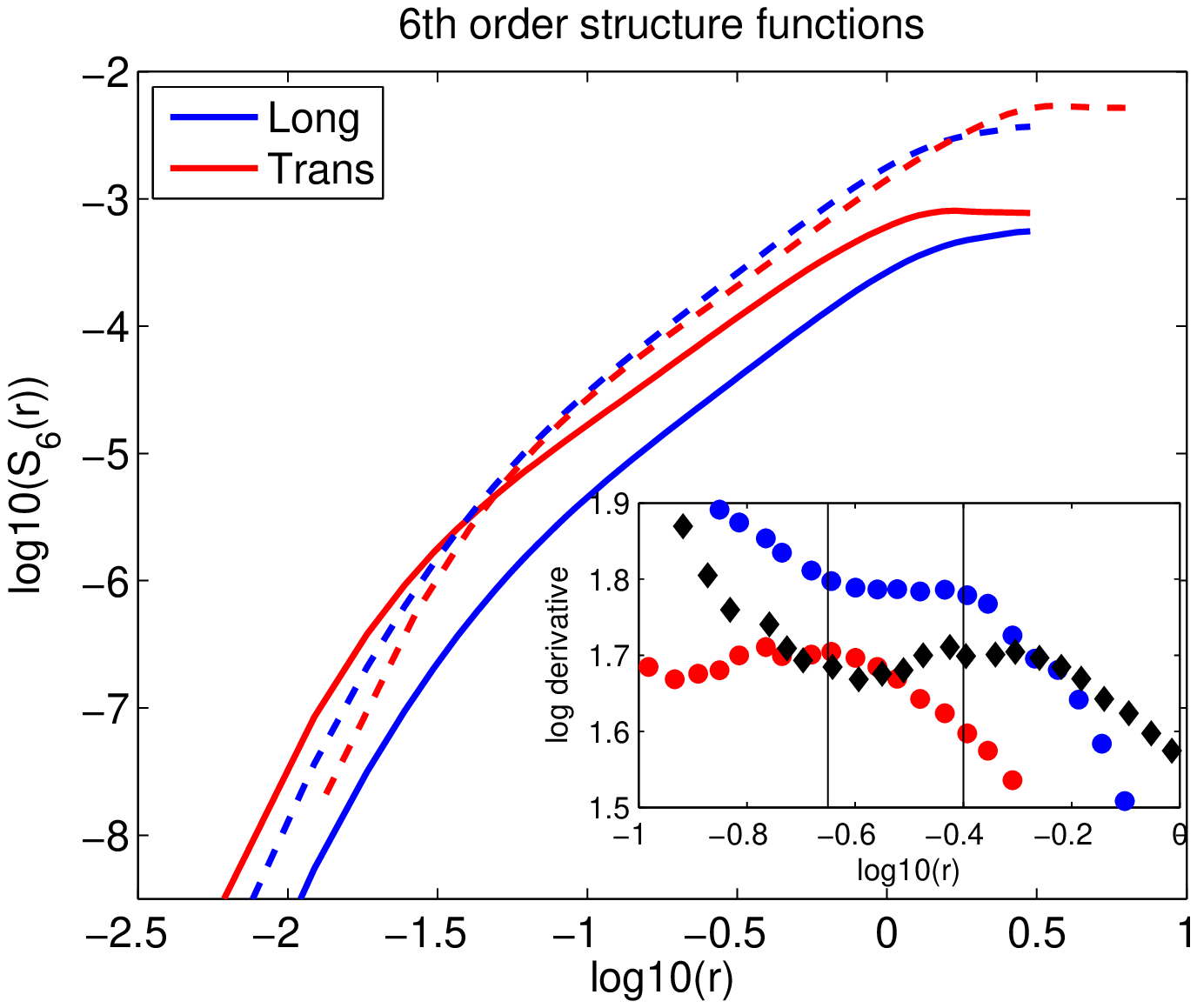} 
  \includegraphics[width=0.45\columnwidth]{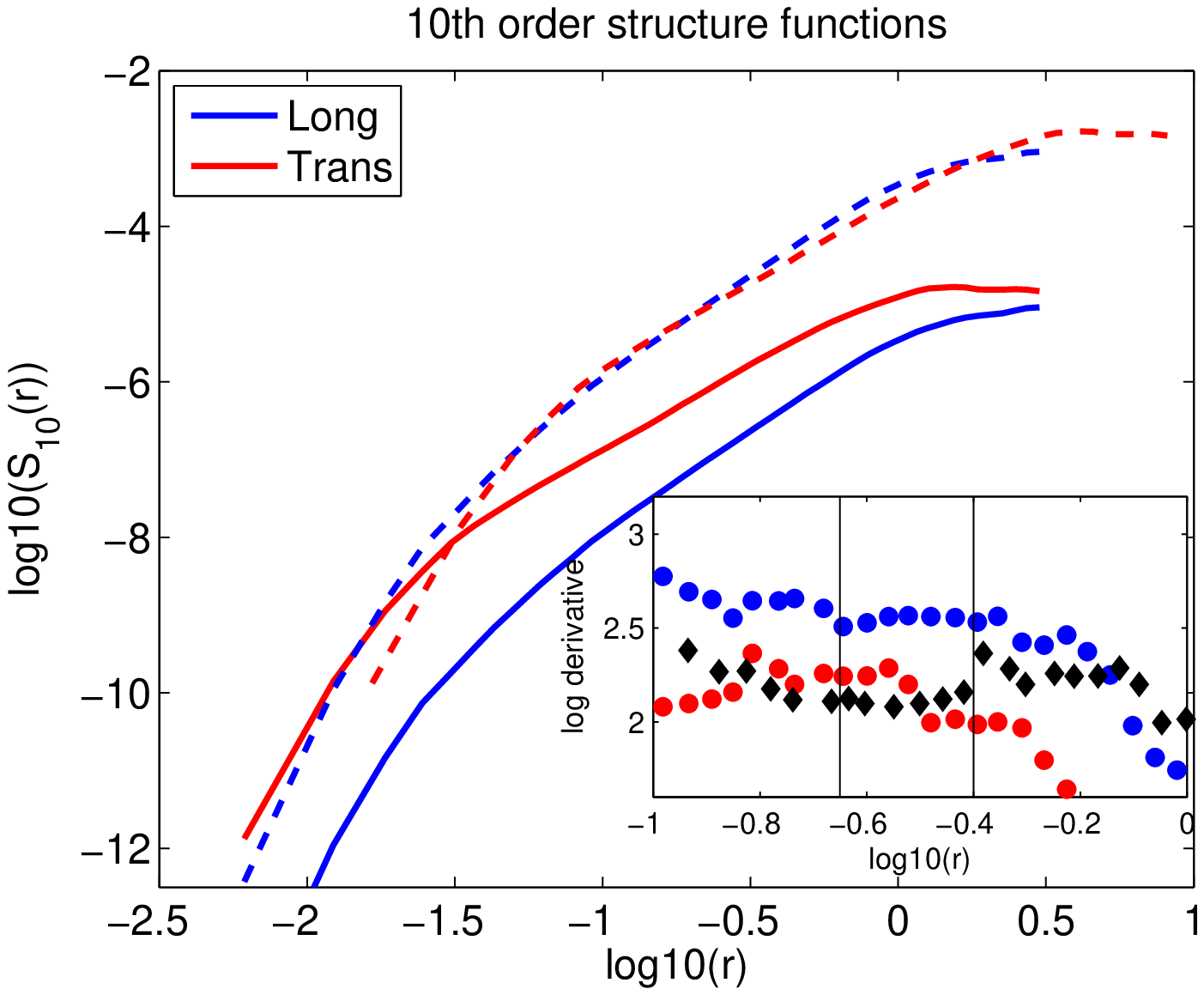}}
  
  \caption{Structure functions of different order ($\Re_\lambda =
    460$). dashed lines: rescaled abscissa according to (6); solid
    lines: original (shifted vertically for clarity), Inset:
    logarithmic derivatives. black diamonds: rescaled transverse
    structure functions. vertical lines indicate the flat region of
    the third order structure function.}
  \label{fig:collaps}
\end{figure*}

Fig.~\ref{fig:collaps} shows the application of the rescaling formula
(\ref{eq:rescaling}) to the 2nd, 4th, 6th and 10th order structure
functions obtained from a Navier-Stokes simulation with $2048^3$ grid
points and parameters as described in Table (\ref{table:param}). We
choose this data set because it contains ten large-eddy turn-over
times and thus provides reliable statistics for high-order structure
functions.  In each sub-figure of Fig. \ref{fig:collaps}, the unscaled
structure functions are shown in the lower part (solid lines) and the
structure functions rescaled according to eqn.(\ref{eq:rescaling}) are
show on top (dashed lines). Note that the original structure functions
are shifted for clarity. The effect of the rescaling-transformation
(\ref{eq:rescaling}) is two-fold: first, the amplitudes from the
dissipative scales up to the integral scales are now very similar. In
addition, the range of scales over which a power-law behavior is
observed is now identical for the two SFs, although the scaling
exponents differ slightly. This is evidence in the inset of each
sub-figure in Fig.~\ref{fig:collaps} where the logarithmic derivatives
of the structure functions with respect to scale have been plotted,
and the vertical lines mark the scaling interval. Note, that both 
effects could not be achieved by an order-independent fixed 
rescaling factor of $3/2$.

\begin{figure}[t]
  \centerline{\includegraphics[width=0.7\columnwidth]{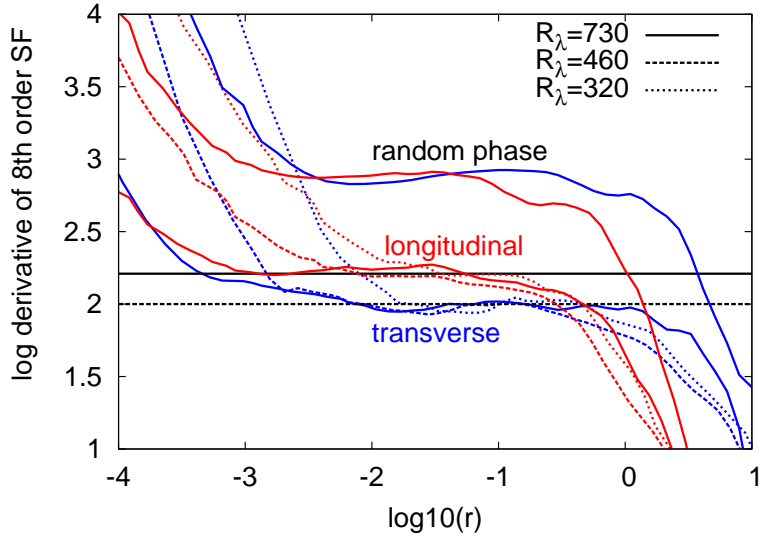}}
  \caption{Logarithmic derivative of longitudinal and
      transverse structure functions for three different Reynolds
      numbers and in the case of a randomized velocity field.}
  \label{fig:newESS}
\end{figure}

For this data set the transverse increments are more intermittent than
the rescaled ones. In order to address the question whether this
difference in scaling might depend on Reynolds number, we show in
Fig.~\ref{fig:newESS} the logarithmic derivative of the 8th order
structure function for three different simulations. With increasing
Reynolds number the inertial range increases but the scaling exponent
(value of the plateau) remains the same. We measure a value of
approx. 2 for the transverse functions and 2.21 for the longitudinal
ones.  For comparison we included data from randomized snapshots
originating from the simulation with $\Re_\lambda = 730$. In detail,
for each Fourier mode we change randomly the phase while preserving
its amplitude and incompressibility of the flow. This preserves the
energy spectrum but destroys the structure of the flow (energy
cascade, coherent structures ...).  The randomized longitudinal and
transverse structure functions exhibit the same scaling exponent, now
close to the trivial 8/3 value. 

A general assumption is that possible remaining large scale anisotropy
in the small scales is expected the decrease with Reynolds number. We
remark that Biferale, Lanotte and Toschi
\cite{biferale-lanotte-etal:2008} showed that the differences in the
high-order exponents remain even if measured in the purely isotropic
sector. That the curves in Fig. \ref{fig:newESS} fall on top of each other within
the inertial range of scales is an indication that the observed
differences in the scaling exponents are not due to large scale
anisotropies. This indicates that the former observed differences have
to be attributed to the specific small scale structures of the
flow. \\

\section{Implications for longitudinal and transverse PDFs}
\begin{figure}[h]
  \centerline{\includegraphics[width=0.9\columnwidth]{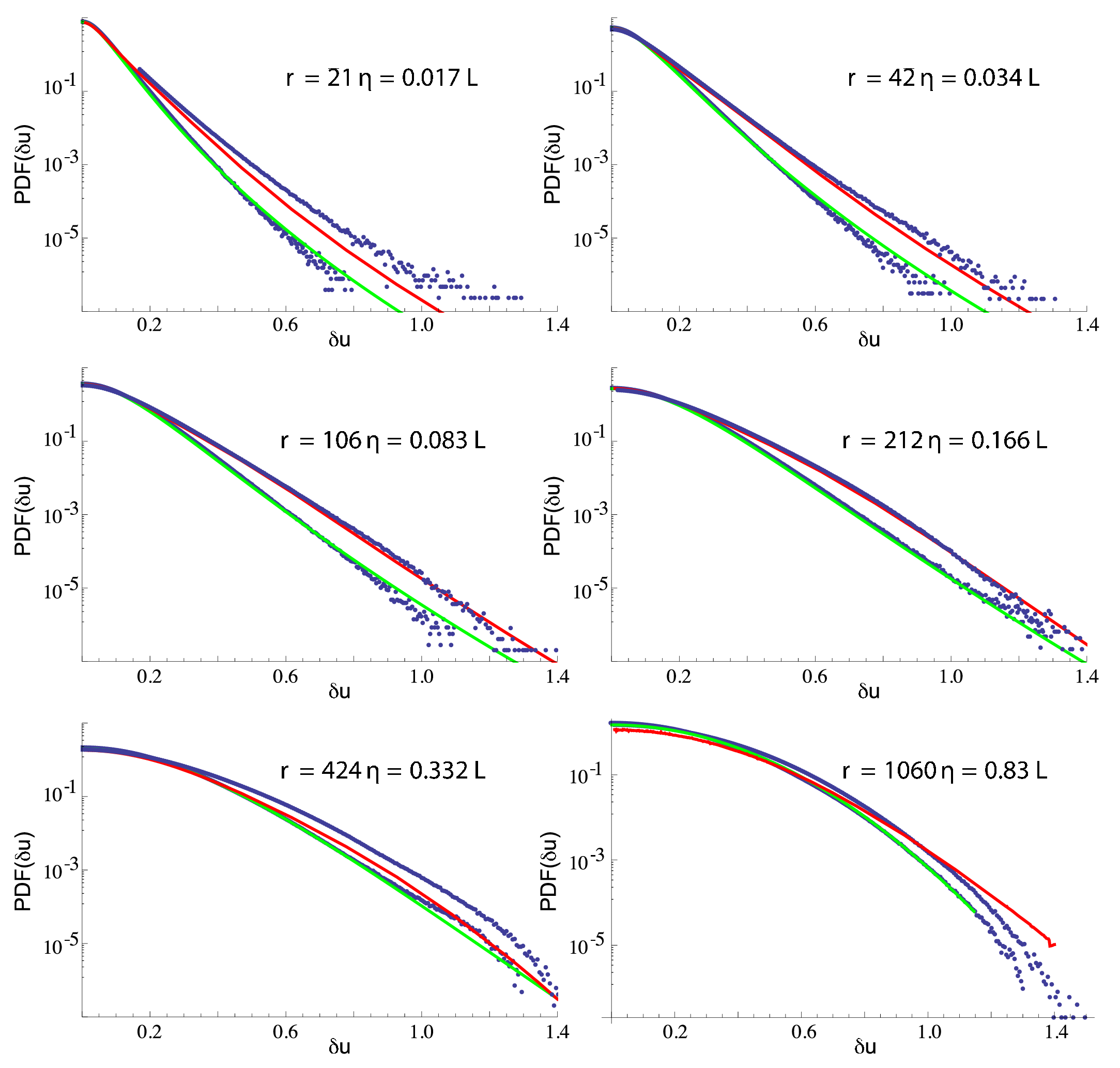}}
  \caption{\label{fig:PDFs} Longitudinal PDF (lower points) and fit
    with log-normal distribution (green line); transverse PDF (upper
    points) and mapped distribution (red line). Shown are the PDFs for
    different increments ranging from the near dissipation range to
    integrals scales. ($\Re_\lambda = 460$, $\eta$ denotes the
  dissipation and $L$ the integral scale.)}
\end{figure}

Since the rescaling property has the effect to make the longitudinal
and transverse structure functions fall nearly on top of each other,
we want to understand the effect of the rescaling transformation on
the probability density functions (PDFs). In this subsection we try to
map longitudinal PDFs to transverse ones using the rescaling property
expressed through eqn.~(\ref{eq:rescaling}). The rescaling
transformations were derived for even order structure functions. Thus
in the following, we disregard skewness effects and consider only the
symmetric part of the PDFs.  We first approximate the numerically
obtained longitudinal PDFs with a log-normal distribution using the
expression given in Yakhot \cite{yakhot:2006}
\begin{equation}
	P_L(\Delta u,r) = \frac{1}{\pi \Delta u \sqrt{\ln r^b}} 
	\int_{-\infty}^\infty e^{-x^2} \exp\left[ 
	-\frac{\left(\ln\frac{\Delta u}{r^a \sqrt{2} x}\right)^2}{4 b \ln r}\right] dx
\end{equation}
for which a fit is obtained with the values $a=0.383$ and $b=0.0166$
\cite{yakhot:2006}. In Fig.~\ref{fig:PDFs} the numerically obtained PDF 
and the fit $P_L(\Delta u,r)$ are shown for several spatial scales.
We apply the inverse Mellin transform
\begin{displaymath}
	P_L(\Delta u , r) = \frac{1}{\Delta u} \int_{-i \infty}^{i \infty} dn \, S(n,r)
	(\Delta u)^{-n}
\end{displaymath}
with $S(n,r) = A(n) r^{\xi(n)}$, and we follow the procedure 
in~\cite{yakhot:2006} which fixes  the amplitude by going to the Gaussian limit
for large spatial differences:
\begin{displaymath}
	A(n) = (n-1)!! = \frac{2^{n/2}}{\sqrt{\pi}}
	\int_{-\infty}^{\infty} e^{-x^2} x^n dx
\end{displaymath}

Now a mapping from the longitudinal PDFs to the transverse PDFs is obtained by
inserting the rescaling relation (\ref{eq:rescaling}) in the expression for the
structure functions:
\begin{equation}
	P_T(\Delta u , r) = \frac{1}{\Delta u} \int_{-i \infty}^{i \infty} dn \, A(n)
	(C(n) r)^{\xi(n)} (\Delta u)^{-n}
	\label{eq:transPDF}
\end{equation}
where $C(n) = \frac{\Gamma(n+2)}{2^n \Gamma^2(n/2+1)}$ as in eqn.
(\ref{eq:rescaling}).

Since the ansatz $S(n,r) = A(n) r^{\xi(n)}$ does not contain a cutoff
at integral and at dissipation scales, this cutoff is inserted in the rescaling
function $C(n)$. In both regions outside the inertial range smooth behavior is
expected. On scales close to the integral range, Gaussian behavior for both
longitudinal and transverse increments is expected and thus no rescaling is
necessary. This justifies to choose $C(n)$ to be constant for $n \le 0$. The
cutoff at the dissipation scale is achieved by choosing $C(n)$ to be constant
for $n > 6$. The precise value of the chosen $n$ is dependent on the actual
Reynolds number and the effect of choosing a different bound allows for a
widening of the transformed PDF. This reflects the fact the Reynolds number has
a similar effect on the width of the PDF.
Evaluating the integral (\ref{eq:transPDF}) using a saddle point
approximation, we obtain a mapping from the log-normal fit of the longitudinal
PDF $P_L(\Delta u, r)$ to a new PDF $P_T(\Delta u, r)$ which is compared to the
numerically obtained data in Fig.~\ref{fig:PDFs} for increments ranging
from the near dissipation range to integrals scales. 
One may observe that the agreement is especially remarkable in the
inertial range ($r = 106 \eta$ and $r =212 \eta$.) We do not expect perfect
agreement for all scales since in that case there would be no room for
differences between longitudinal and transverse structure functions. Thus the
discrepancy for $r = 21 \eta$ and $r = 42 \eta$ just represents the missing
contributions of the pressure term.
Therefore, this method of mapping the PDFs is also a promising
candidate for applications like PDF modeling of turbulent flows (see
Pope~\cite{pope:2000} and references therein).

\section{Conclusions and outlook} 
In this paper, we have suggested a new way of analyzing experimental and
numerical data for longitudinal and transverse structure functions in Eulerian
data of a turbulent velocity field. This procedure yields a mapping between the
longitudinal and transverse scales, which provides consistent reference point in the 
identification of the inertial ranges of scales of turbulent flows. In addition, the derived  
scale correspondence allows for a direct mapping of the {\it full} probability density of 
transverse and longitudinal structure functions. This may be of much practical interest as the 
distributions carry a more complete information that than a subset of their moments.
The gap of longitudinal and transverse structure function exponents seems not to
depend on Reynolds number but on small scale structure of the flow. The
proposed mapping may help clarify the role played by the pressure terms. Future
work will be devoted to the analysis of other turbulent systems like
magnetohydrodynamics, where the addition of the magnetic pressure term poses an
interesting comparison. \\

\noindent {\bf Acknowledgments} 
Access to the IBM BlueGene/P computer JUGENE at the FZ J\"ulich was made
available through the 'XXL-project' of HBO28. This work benefited from partial
support through DFG-FOR1048 in Germany, and ANR-07-BLAN-0155 in France.

\section*{References}

\bibliographystyle{iopart-num.bst}
\bibliography{bib}

\end{document}